\documentclass[twocolumn,pra,nofootinbib,superscriptaddress,aps,floatfix]{revtex4}
\usepackage{graphicx}
\usepackage{bm}
\usepackage{amsmath}
\usepackage{color}

\begin{document}
\title{\bf{Dynamical normal modes for time-dependent Hamiltonians in two dimensions}}
\author{I. Lizuain}
\affiliation{Department of Applied Mathematics, University of the Basque Country UPV/EHU, Plaza Europa 1, 20018 Donostia-San Sebastian, Spain}
\author{M. Palmero}
\affiliation{Departament of Physical Chemistry, University of the Basque Country UPV/EHU, Apdo. 644, Bilbao 48080, Spain}
\author{J. G. Muga}
\affiliation{Departament of Physical Chemistry, University of the Basque Country UPV/EHU, Apdo. 644, Bilbao 48080, Spain}
\begin{abstract}
We present the theory of time-dependent point transformations to find independent dynamical normal modes for  2D systems subjected to  time-dependent control in the limit of small oscillations. The condition that determines if the independent modes can indeed be defined 
is identified, and a geometrical analogy is put forward. The results explain and unify recent work to design fast operations on trapped ions,
needed to implement a scalable quantum-information architecture: transport, expansions, and the  separation of two ions, two-ion phase gates, as well as the rotation of an anisotropic trap for an ion are treated and shown to be analogous to a mechanical system of two masses connected by springs with time dependent stiffness.          
\end{abstract}
\maketitle

\section{Introduction}
The small oscillations regime of systems composed by interacting particles is best characterized,   
and possibly controlled \cite{Cirac}, using a decomposition of the dynamics into independent normal modes. 
They are concerted and harmonic motions of all particles in the system with frequencies 
that may be found by diagonalizing the harmonic part of the potential around the equilibrium,
using a point canonical transformation that also defines the normal-mode coordinates.    
The analysis is usually made for  time-independent interaction potentials, see e.g. \cite{Steane, James, Mori,Hom}
in the context of trapped ions, 
but the potential may in principle be also modified externally 
in a time dependent manner. Generalizing the normal modes for these
time-dependent scenarios is necessary  as our ability to drive microscopic or macroscopic systems  
improves with technological advances, see e.g \cite{Landa,Landa2}.         
In this paper we study the possibility to define independent dynamical normal modes 
in systems described by two dimensional, time-dependent Hamiltonians.  
While the question is interesting per se and relevant for a broad span of externally controllable physical systems 
near equilibrium,   
our main motivation has been the need to understand and possibly improve on 
recent work to inverse-engineering fast and robust operations to drive the motion of trapped ions \cite{PTGM2013,PBGLM2014,PMAHM2015,PMPRM2015,LPRCM2015,PWGSM2016,phaseg}. 
Trapped ions  constitute the most developed physical platform to  
implement quantum information processing. Since many ions 
in a single trap are difficult to control, a route towards  
large scale computations with many qubits relies on a ``divide and conquer'' scheme \cite{Wine1,Wine2},    
where ions are shuttled around in multisegmented Paul traps that hold just a few ions in each processing site.  
Apart from shuttling, complementary operations such as separating and merging ion chains, rotations,  and expansions/compressions
may be needed.  Coulomb interactions, and controllable external effective potentials determine the Hamiltonians that govern the motion of the ions,     
approximated in quadratic form near equilibrium.  
Independent dynamical normal modes (``dynamical'' because their definition depends on time due to the changing external 
control) are very useful, not only to describe the motion in a simple way, but also to inverse-engineer the dynamical operations.   
Specifically, for operations on one ion in a two-dimensional (2D) potential
or on two ions interacting in  a one-dimensional (1D) line\cite{PTGM2013,PBGLM2014,PMAHM2015,PMPRM2015,LPRCM2015,PWGSM2016,phaseg}, we noticed that uncoupled dynamical normal modes 
cannot always be defined. Each case was analyzed  separately but  
a generic understanding of the conditions that determine the coupling/uncoupling of normal-mode coordinates was missing.    
This paper presents first in Sec. \ref{the_model_sec} a  comprehensive theory where the criterion for separability into independent
motions by time-dependent point transformations is identified. 
In Sec. \ref{ions_section} the theory and criterion are applied to different operations on trapped ion systems. 
After a final discussion and outlook for future work, 
Appendix \ref{spring_system_appendix} shows that the general Hamiltonian structure considered describes a mechanical model of two masses connected to walls and to each other by  springs with time-dependent stiffness.     
The treatment in the main text is classical but the results are also valid in the quantum domain, 
as shown in  Appendix \ref{quantum_appendix}.           
%

\section{The model}
\label{the_model_sec}
Our starting point is a 
2D Hamiltonian for two interacting particles moving on a line, with masses $m_1$ and $m_2$,   
(1D) coordinates $q_1$, $q_2$,  conjugate momenta $p_1$, $p_2$, and time-dependent potential $U(q_1,q_2;t)$,    
\begin{eqnarray}
H=\frac{p_{1}^2}{2m_1}+\frac{p_{2}^2}{2m_2}+U(q_1,q_2;t).
\end{eqnarray}
The same Hamiltonian structure, with $m=m_1=m_2$, may also describe one particle 
moving on a two dimensional surface with potential $U$.  
The first step is to find the equilibrium positions, $q_1^{(0)}$,  $q_2^{(0)}$, from the potential minimum 
given by $\nabla U=0$, and expand $U$ at that point retaining only quadratic terms,      
\begin{eqnarray}
\label{starting_Hqp}
H&=&\frac{p_{1}^2}{2m_1}+\frac{p_{2}^2}{2m_2}
\nonumber\\
&+&\frac{1}{2}\sum_{i,j=1}^2 k_{ij}[q_i-q_i^{(0)}(t)][q_j-q_j^{(0)}(t)].
\end{eqnarray}
%
Because of the generic time dependence of $U$,  the coeffiecients $k_{ij}(t)=\frac{\partial^2 U}{\partial q_i \partial q_j}\Big|_{{q_1^{(0)},q_2^{(0)}}}$
and the equilibrium positions $q_i^{(0)}(t)$ may depend on time, 
but the explicit time dependence will generally be omitted hereafter to avoid a cumbersome notation.
The $k_{ij}$ coefficients in Eq. (\ref{starting_Hqp}) are the elements of the real and symmetric  
$2\times 2$ matrix $K$. 
Being symmetric, it may be parameterized as 
\begin{eqnarray}
\label{Kgen}
 K&=&\left(  \begin{matrix}  k+k_1&-k\\-k&k+k_2  \end{matrix}\right),
\end{eqnarray}
where $k,k_1,k_2$ are generally time dependent.  If they are positive, $K$ is a positive matrix (with positive eigenvalues). 
Defining now the vector 
\begin{equation}
\label{psi_tr}
\psi^T=\left(q_1-q_1^{(0)},q_2-q_2^{(0)},p_{1},p_{2}\right),
\end{equation}
and the mass matrix $M$,
\begin{equation}
\label{mass_matrix}
 M=\left(\begin{matrix}m_1&0\\0&m_2   \end{matrix}\right),
\end{equation}
the  Hamiltonian (\ref{starting_Hqp}) can be written in a compact matrix representation as
\begin{equation}
\label{starting_H}
H=\frac{1}{2}\psi^T W \psi,
\end{equation}
where $W$ is the $4\times 4$ symmetric matrix formed by $K$ and $M^{-1}$ $2\times 2$ blocks, 
\begin{equation}
W=\left(  \begin{matrix}  K&0\\ 0&M^{-1}  \end{matrix}\right).
\end{equation}
Interestingly, the Hamiltonian (\ref{starting_H}) corresponds as well to a system of two  masses connected to walls and to each other 
by time-dependent spring constants, 
see Appendix \ref{spring_system_appendix}.

The main goal of this paper is to investigate if there is a point transformation producing new coordinates 
$Q_1, Q_2$ and momenta $P_1,P_2$ such that the corresponding Hamiltonian $\widetilde H(Q_1,Q_2,P_1,P_2)$ does not have cross terms and can
be separated into  independent harmonic motions.    
We shall see that this is not always possible and we will give the conditions to be satisfied by $H$ in order to successfully 
separate $\widetilde H$ by a time-dependent point transformation. Some alternative treatments when the decomposition fails will also be pointed out. 

%
%
%
%
%
\subsection{Time dependent point canonical transformation}
Let us consider the general time-dependent (linear) change of coordinates  
\begin{equation}
\label{A_change_coordinates}
\left(  \begin{matrix}  Q_1\\Q_2   \end{matrix}\right)
=A(t)\left(  \begin{matrix}  q_1-q_1^{(0)}\\q_2-q_2^{(0)}   \end{matrix}\right),
\end{equation}
where $A(t)$ is a $2\times2$ matrix to be determined, invertible at all times.
%
%
This transformation is generated by the type-$2$ generating function \cite{Goldstein}
 \begin{eqnarray}
   \label{f2_generating}
  F_2&=&\sum_{i=1}^2 P_i Q_i(q_1,q_2)=(P_1,P_2)A(t)\left(  \begin{matrix} q_1-q_1^{(0)}\\q_2-q_2^{(0)}     \end{matrix}\right).
  \nonumber\\
 \end{eqnarray}
The momenta transform according to $p_i=\partial_{q_i} F_2$,
\begin{eqnarray}
   p_1&=&\frac{\partial F_2}{\partial q_1}=a_{11}P_1  + a_{21}P_2,\nonumber\\
   p_2&=&\frac{\partial F_2}{\partial q_2}=a_{12}P_1  + a_{22}P_2,\nonumber
\end{eqnarray}
which can be written in matrix form as
\begin{equation}
\left(  \begin{matrix}  p_1\\p_2   \end{matrix}\right)
=A^T\left(  \begin{matrix}  P_1\\P_2   \end{matrix}\right),
\label{pP}
\end{equation}
where $A^T$ denotes the transpose of $A$. 
In the $4$-dimensional representation introduced previously, the canonical transformation of 
coordinates and momenta is compactly given by
\begin{equation}
\label{transf_direct}
 \widetilde \psi=
 \left(  \begin{matrix}  A&0\\0&A^{-T} \end{matrix}\right)\psi,
\end{equation}
where $\psi^T=\left(q_1-q_1^{(0)},q_2-q_2^{(0)},p_1,p_2\right)$, $\widetilde{\psi}^T=(Q_1,Q_2,P_1,P_2)$, and $A^{-T}\equiv(A^T)^{-1}$ stands for the 
inverse of the transpose of $A$.
%
%
%
%
\subsection{Inertial effects and effective Hamiltonian}

As a consequence of the time dependence of the potential, 
the coordinate transformation may correspond to a description in a non-inertial frame, 
where inertial forces appear.    
The transformed Hamiltonian in the new coordinates will read $\widetilde H=H+\partial_t F_2$,
where the last term accounts for inertial effects arising due to the 
explicit time dependence of $F_2$, 
    \begin{eqnarray}
     \frac{\partial F_2}{\partial t}&=&(P_1,P_2)\dot A \left(  \begin{matrix}  q_1-q_1^{(0)}\\  q_2-q_2^{(0)}   \end{matrix}\right)
     - (P_1,P_2) A \left(  \begin{matrix}   \dot q_1^{(0)}\\  \dot q_2^{(0)}   \end{matrix}\right)
   \nonumber\\
     &=&   (P_1,P_2)\dot A A^{-1} \left(  \begin{matrix}  Q_1\\  Q_2   \end{matrix}\right)
     - (P_1,P_2) A \left(  \begin{matrix}  \dot q_1^{(0)}\\  \dot q_2^{(0)}   \end{matrix}\right)\!\!,
     \label{non_inertial}
    \end{eqnarray}
and  the dots denote time derivatives. The inertial effects have two different contributions, 
a quadratic term proportional to $\dot A$, and a linear term proportional to $\dot q_i^{(0)}$.

%
%
%
%

Using the coordinate and momenta transformations (\ref{transf_direct}) and the inertial terms (\ref{non_inertial}), 
the transformed Hamiltonian in the new coordinates can be written as
\begin{eqnarray}
\widetilde H&=&H+\frac{\partial F_2}{\partial t}\nonumber\\
&=&\frac{1}{2}\tilde \psi^T \widetilde W \tilde \psi
-(P_1,P_2) A \left(  \begin{matrix}  \dot q_1^{(0)}\\  \dot q_2^{(0)}   \end{matrix}\right),
\label{tilde_H}
 \end{eqnarray}
with
\begin{eqnarray}
  \widetilde W&=&
  \left( \begin{matrix}  A^{-T}KA^{-1}&(\dot A A^{-1})^T\\\dot A A^{-1}&AM^{-1}A^T  \end{matrix}\right).
\label{tilde_W}
 \end{eqnarray}
Our aim now is to find a transformation matrix $A$ such that $\widetilde W$ is a diagonal $4\times 4$ matrix.  
Since the linear part in the Hamiltonian  (\ref{tilde_H}) is already uncoupled, 
this would define ``dynamical normal modes'' \cite{PBGLM2014}, evolving independently of each other.

%
%
\subsection{Diagonalization of $\widetilde H$}
To have an uncoupled effective Hamiltonian $\widetilde H$ (diagonal $\widetilde W$),
two conditions have to be satisfied: the diagonal blocks in Eq. (\ref{tilde_W}) have to be diagonal $2\times 2$ matrices,
and the off-diagonal blocks  should vanish for all times.
 
The first one amounts to simultaneously diagonalizing two  bilinear forms \cite{Goldstein}.
As  the masses are positive quantities, the square root of the matrix $M$ given in Eq. (\ref{mass_matrix}) 
can be defined as
\begin{eqnarray}
M^{1/2}&=&\textrm{diag}\left(\sqrt{m_1},\sqrt{m_2}\right).
\end{eqnarray}
We now define the ``mass-weighted potential'' as
\begin{eqnarray}
\label{mass_weighted_K}
 \widetilde K=M^{-1/2} K  M^{-1/2},
\end{eqnarray}
which 
is also symmetric since $K$ is symmetric.
The explicit expression of the  mass-weighted potential $\widetilde K$ is
\begin{eqnarray}
\label{tilde_K_springs}
 \widetilde K&=&M^{-1/2} K M^{-1/2}=
 \left(  \begin{matrix}  \frac{k+k_1}{m_1}&\frac{-k}{\sqrt{m_1m_2}}\\\frac{-k}{\sqrt{m_1m_2}}&\frac{k+k_2}{m_2}  \end{matrix}\right)\!,
\end{eqnarray}
which, for positive masses, is also positive definite if $K$ is positive. 
Since $\widetilde K$ is in any case symmetric, it can be diagonalized by means
of an orthogonal matrix $\mathcal{O}$,
\begin{eqnarray}
\mathcal{O}^T \widetilde K \mathcal{O}=\textrm{diag}(\Omega_1^2,\Omega_2^2),
\end{eqnarray}
with
\begin{equation}
\label{O_general}
 \mathcal{O}=\left(\begin{matrix} \cos\theta&-\sin\theta\\ \sin\theta&\cos\theta \end{matrix}\right),
\end{equation}
and where the  time-dependent parameter $\theta$ is given by the relation
\begin{equation}
\label{tan2theta}
\tan 2\theta=\frac{2k\sqrt{m_1m_2}}{m_1(k+k_2)-m_2(k+k_1)}.
\end{equation}
$\Omega_i^2$, the eigenvalues of $\widetilde K$, give the time-dependent eigenfrequencies of each normal mode, with explicit expressions
\begin{eqnarray}
 \Omega_1^2&=&\left(\frac{k+k_1}{m_1}\right)\cos^2\theta + \left(\frac{k+k_2}{m_2}\right)\sin^2\theta
 \nonumber\\
 &-& \frac{k}{\sqrt{m_1m_2}}\sin 2\theta,
 \nonumber\\
 \Omega_2^2&=&\left(\frac{k+k_1}{m_1}\right)\sin^2\theta + \left(\frac{k+k_2}{m_2}\right)\cos^2\theta
 \nonumber\\ 
 &+& \frac{k}{\sqrt{m_1m_2}}\sin 2\theta,
 \label{eigenvalues_K_tilde}
\end{eqnarray}
positive if $k$, $k_1$, $k_2$ are all positive. 
The ``modal matrix''  
\begin{equation}
\label{A_expl_gral}
A=\mathcal{O}^TM^{1/2}=\left(  \begin{matrix} \sqrt{m_1} \cos\theta &\sqrt{m_2}\sin\theta \\  -\sqrt{m_1} \sin\theta &\sqrt{m_2}\cos\theta \end{matrix} \right)
\end{equation}
diagonalizes simultaneously both the blocks with $M^{-1}$ and $K$ in the main diagonal of Eq. (\ref{tilde_W}) since
\begin{eqnarray}
 A^{-T}KA^{-1}&=&\textrm{diag}(\Omega_1^2,\Omega_2^2),
 \\
 AM^{-1}A^T&=&1.
\end{eqnarray}
Normal mode coordinates  $\{Q_1,Q_2\}$ are defined by 
the transformation (\ref{A_change_coordinates}) with $A$ given by  Eq. (\ref{A_expl_gral}),
 \begin{eqnarray}
  \label{full_transf}
\left(  \begin{matrix}  Q_1\\Q_2   \end{matrix}\right)
=\left(  \begin{matrix} \sqrt{m_1} \cos\theta &\sqrt{m_2}\sin\theta \\  -\sqrt{m_1} \sin\theta &\sqrt{m_2}\cos\theta \end{matrix} \right)
\left(  \begin{matrix}  q_1-q_1^{(0)}\\q_2-q_2^{(0)}   \end{matrix}\right).
 \end{eqnarray}
Note that we have not proved yet if  they are uncoupled. They will  be independent if the 
non-diagonal term $\dot A A^{-1}$ in Eq. (\ref{tilde_W}) vanishes.  With the explicit expression of $A$ in (\ref{A_expl_gral}) we can  calculate  the
$ \dot AA^{-1}$ term, 
\begin{equation}
\label{coupling_term_V0}
 \dot AA^{-1}=\dot\theta\left(  \begin{matrix}  0&1\\-1&0   \end{matrix}\right).
\end{equation}
Including this coupling, the effective Hamiltonian takes the form 
 \begin{eqnarray}
 \widetilde H&=&\frac{1}{2}\sum_{i=1}^2\left(P_i^2+\Omega_i^2Q_i^2\right)
 -(P_1,P_2) A \left(  \begin{matrix}  \dot q_1^{(0)}\\  \dot q_2^{(0)}   \end{matrix}\right)-\dot\theta L_z,
   \nonumber\\
 \label{H_tilde_general}
 \end{eqnarray}
where $L_z=Q_1P_2-Q_2P_1$ has the form of the $z$-component of an angular momentum.
%
%

We conclude that if  $\theta$ does not depend on time, the modes are uncoupled. 
As we shall see in several examples,  some configurations of  the matrices ${K}$ and $M$  
lead to $\dot{\theta}=0$, even if the $k$, $k_1$, $k_2$ are time dependent: for example  
$k_1=c_1k, k_2=c_2k$  with constants $c_1,c_2$;  $k_1=k_2$ for $m_1=m_2$; or  $k=0$, with time dependent $k_1$ and $k_2$.     
If $\theta$ is time independent,
the new coordinates define indeed independent ``dynamical normal modes''
with an uncoupled Hamiltonian
 \begin{eqnarray}
  \widetilde H&=&\frac{1}{2}\sum_{i=1}^2\left(P_i^2+\Omega_i^2Q_i^2\right)
  - (P_1,P_2) A \left(  \begin{matrix}  \dot q_1^{(0)}\\  \dot q_2^{(0)}   \end{matrix}\right).
  \label{H_tilde_gral}
 \end{eqnarray}
 At this point it is customary to perform a momentum shift, so that the new Hamiltonian 
 includes a term linear in coordinates rather than a term linear in momentum. This is done with the 
 generating function 
 \begin{equation}
 F_2=\sum_{i=1}^2 (P'_i+P_{0,i})Q_i,\nonumber
 \end{equation}
 where 
 \begin{equation}
 \left(\begin{matrix} P_{0,1}\\P_{0,2}\end{matrix}\right)\equiv A \left(  \begin{matrix}  \dot q_1^{(0)}\\  \dot q_2^{(0)}   \end{matrix}\right),\nonumber
 \end{equation}
 which gives for the new momenta and coordinates
 \begin{eqnarray}
 P'_i&=&P_i-P_{0,i},
 \nonumber\\
 Q'_i&=&Q_i,
\nonumber
 \end{eqnarray}
 and $\partial_t F_2= \sum_{i=1}^2  \dot{P}_{0,i} Q'_i$, so that the transformed Hamiltonian, up to 
 purely time dependent terms that can be added or subtracted without changing the physics, 
 takes the form of two moving harmonic oscillators,
 \begin{equation} 
\widetilde{H}'=\frac{1}{2}\sum_{i=1}^2 \left[  {P'}_i^2+\Omega_i^2\left({Q'}_i+\frac{\dot{P}_{0,i}}{\Omega_i^2}\right)^2\right].\nonumber
 \end{equation}
While this is the form that has been used to speed up several operations on trapped ions  \cite{PTGM2013,PBGLM2014,PMAHM2015,PMPRM2015,LPRCM2015,PWGSM2016,phaseg}, for the discussion of
the separability of these systems in Sec. \ref{ions_section} it is enough to examine $\tilde{H}$ and we shall omit 
the momentum shift transformation.  
%

\subsection{Geometrical Interpretation}
\begin{figure}
\includegraphics[width=7.5cm]{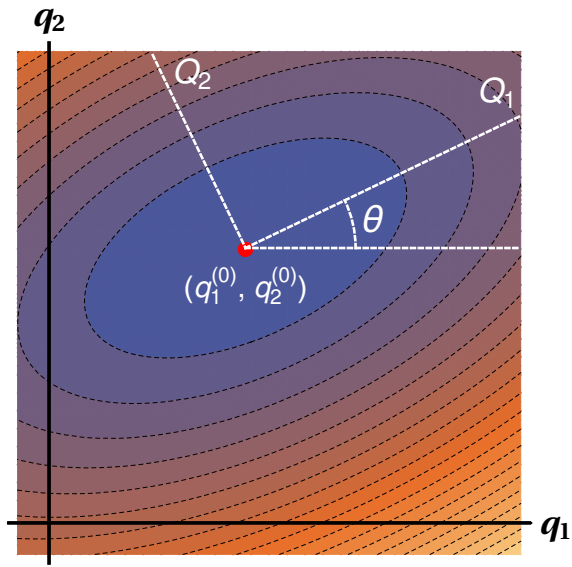}
\caption{(Color online) Schematic representation of iso-potential curves of a mass-weighted potential in the 2-dimensional
configuration space of lab-frame coordinates $\{q_1,q_2\}$. 
These curves are ellipses centered at the moving equilibrium position $\left(q_1^{(0)},q_2^{(0)}\right)$
with  the orientation of the principal axes given by the angle $\theta$.
The dynamical normal  mode coordinates $\{Q_1,Q_2\}$ are translated from the origin but also rotated by $\theta$. }
\label{ellipse_fig}
\end{figure}


We  have just shown that a condition  to  define independent modes  by a point transformation 
is that  $\theta$ (and therefore the transformation $A$) does not depend on time. What does this parameter represent?

Let us now visualize the mass-weighted symmetric potential $\widetilde K$ in Eq. (\ref{tilde_K_springs})
as a matrix defining a quadratic form. Quadratic forms are geometrically represented by conic sections.
If  $\widetilde K$ is positive definite (i. e., with positive eigenvalues), the conic section  
defined by $\widetilde K$ is an ellipse centered at the moving equilibrium position $\left(q_1^{(0)},q_2^{(0)}\right)$.
These ellipses are  iso-potential curves of the mass-weighted potential $\widetilde K$ in the $2$-dimensional configuration space 
$\lbrace{q_1,q_2\rbrace}$, see Fig. {\ref{ellipse_fig}. 
The principal axes theorem states that the orthonormal coordinate system where the ellipse is well-oriented is given by
the orthonormal eigenvectors of $\widetilde K$, 
while the inverse of the square root of its eigenvalues are the radii of the corresponding axes. 
The orthogonal matrix (\ref{O_general}) is formed by the eigenvectors of $\widetilde K$,
\begin{eqnarray}
 v_1^T&=&(\cos\theta,\sin\theta)\nonumber
 \\
 v_2^T&=&(-\sin\theta,\cos\theta).\nonumber
\end{eqnarray}
These vectors define the orthonormal coordinate system where the ellipse is well-oriented, see Fig. \ref{ellipse_fig}. Therefore the parameter $\theta$ 
gives us the orientation of the ellipse. 
More generally, it gives the orientation of the principal axes
if $\widetilde{K}$ is not positive (The engineering of fast dynamics may require that an eigenfrequency 
becomes transiently an imaginary number \cite{ChenPRL}).   
In general for a time dependent potential, the equilibrium position, 
the shape (size of principal axes) and orientation (angle $\theta$) will vary in time, 
but only the rotation of the principal axes couples the normal modes. 
%

\section{Application to trapped-ion systems}
\label{ions_section}
In this section, we apply the results of the previous sections to systems of two  ions in a linear trap, or one ion in a two dimensional trap. 
In particular we analyze if independent 
dynamical normal modes can be defined by point transformations.  
This is a key issue  to design and engineer fast and robust protocols for operations needed 
to develop scalable quantum information
 processing:  ion transport \cite{PTGM2013,PBGLM2014,LPRCM2015}, 
trap expansions or compressions \cite{PMAHM2015}, ion splitting \cite{PMPRM2015},
 phase gates \cite{phaseg}, or rotations \cite{PWGSM2016}.

\subsection{Transport and expansions (or compressions) of two interacting trapped ions}
%
Let us consider two singly-charged, positive ions in a linear trap at lab-frame coordinates $q_1$ and $q_2$,  
coupled via Coulomb interaction and trapped in an external, possibly time dependent, harmonic potential \cite{PBGLM2014,PMAHM2015}.
The external potential can be translated \cite{PBGLM2014} 
and, in addition, expanded or compressed \cite{PMAHM2015}.
The Hamiltonian describing this system  is
\begin{eqnarray}
 H&=&\frac{p_1^2}{2m_1}+\frac{p_2^2}{2m_2}+U,
 \nonumber\\
  U&=&\frac{1}{2}\sum_{i=1}^2k(t)\left[q_i-Q_0(t)\right]^2
 +\frac{C_c}{q_1-q_2}.
 \label{H_ion_transport_lab}
 \end{eqnarray}
Here, $C_c=\frac{e^2}{4\pi\epsilon_0}$ is the Coulomb constant
and $k(t)$ is the common (time-dependent) spring constant that determines  the oscillation frequency of 
each ion ($\omega_i^2=k/ m_i$) in the absence of Coulomb coupling \cite{PMAHM2015}.
$Q_0=Q_0(t)$ defines the position of the minimum of the external potential, i. e., the position of the, possibly moving, trap \cite{PBGLM2014}.
We can also set $q_1>q_2$ because of the strong Coulomb repulsion \cite{PTGM2013}.
If the ions are sufficiently cold, they ``crystallize'' around the classical equilibrium positions $q_i^{(0)}$, which are solutions of the set of equations 
$\frac{\partial U}{\partial q_i}=0$ for $i=1,2$, 
\begin{eqnarray}
 q_1^{(0)}&=&Q_0+\left(\frac{C_c}{4k}\right)^{1/3}=Q_0+\frac{q_0}{2},\nonumber\\
 q_2^{(0)}&=&Q_0-\left(\frac{C_c}{4k}\right)^{1/3}=Q_0-\frac{q_0}{2},\nonumber
\end{eqnarray}
where $q_0=q_1^{(0)}-q_2^{(0)}=(2C_c/k)^{1/3}$ is the equilibrium distance between ions\footnote{We omit the explicit 
time dependence of $k=k(t)$ and $Q_0=Q_0(t)$ hereafter to avoid a cumbersome notation}.
These positions are time dependent but independent of the mass, as we assume that the external potential
is due to trap electrodes
that interact only with the ionic charge.
%
If we now approximate the coupling potential $U$ by its Taylor expansion truncated to second
order in $q_i-q_i^{(0)}$ (small displacements from equilibrium), 
we end up with a quadratic potential.
Therefore, up to a purely time dependent term, we can approximate the  Hamiltonian (\ref{H_ion_transport_lab}) harmonically  as
\begin{eqnarray}
 H&=&\frac{p_1^2}{2m_1}+\frac{p_2^2}{2m_2}\nonumber\\
 &+&\frac{1}{2}\!\left(q_1\!-\!q_1^{(0)},q_2\!-\!q_2^{(0)}\right)
K
 \left(  \begin{matrix}  q_1\!-\!q_1^{(0)}\\  q_2\!-\!q_2^{(0)}  \end{matrix}\right)
 \label{H_transport_harmonic}
 \end{eqnarray}
 with the $K$ matrix given by
 \begin{equation}
  K=\left(  \begin{matrix}  2k&-k\\  -k&2k  \end{matrix}\right).\nonumber
 \end{equation}
%
This Hamiltonian corresponds to the case $k_1=k_2=k$, so that 
Eq. (\ref{tan2theta}) gives a constant $\theta$ even for a time-dependent $k$, 
\begin{eqnarray}
\label{tan2theta_equal_k}
\tan2\theta&=&\frac{\sqrt{m_1m_2}}{m_1-m_2}.
\end{eqnarray}
The normal modes are  given by (\ref{full_transf})
\begin{eqnarray}
\!\!\left(\!  \begin{matrix}  Q_1\\Q_2   \end{matrix}\!\right)
=\left(\!  \begin{matrix} \sqrt{m_1} \cos\theta &\sqrt{m_2}\sin\theta \\  -\sqrt{m_1} \sin\theta &\sqrt{m_2}\cos\theta \end{matrix} \right)\!
\left(\!  \begin{matrix}  q_1\!-\!\left(Q_0+q_0/2\right)\\q_2\!-\!\left(Q_0-q_0/2\right)\end{matrix}\right)\!,\nonumber
\label{dyn_normal_modes_transport}
\end{eqnarray}
with $\theta$ given by relation (\ref{tan2theta_equal_k}). These normal modes depend on time through the time dependent parameters $Q_0$ and $q_0$.
In these coordinates, the Hamiltonian (\ref{H_transport_harmonic}) 
transforms to the diagonal and uncoupled form (\ref{H_tilde_gral}),
\begin{eqnarray}
  \widetilde H&=&\frac{1}{2}\sum_{i=1}^2\left(P_i^2+\Omega_i^2Q_i^2\right)
  - (P_1,P_2) A \left(  \begin{matrix}  \dot Q_0+\dot q_0/2\\  \dot Q_0-\dot q_0/2  \end{matrix}\right),\nonumber\\
  \label{H_tilde_transport}
 \end{eqnarray}
where 
%
 \begin{eqnarray}
 \Omega_1^2&=&k\left(\frac{2\cos^2\theta}{m_1} + \frac{2\sin^2\theta}{m_2} -\frac{\sin 2\theta}{\sqrt{m_1m_2}}\right),\nonumber\\
 \Omega_2^2&=&k\left(\frac{2\sin^2\theta}{m_1} + \frac{2\cos^2\theta}{m_2} +\frac{\sin 2\theta}{\sqrt{m_1m_2}}\right),
 \end{eqnarray}
which may depend on time because of $k=k(t)$. 
Note that the inertial effects in this Hamiltonian (\ref{H_tilde_transport}) are only included into the linear-in-momentum term
and are due to the transport of the trap ($\dot Q_0$ term) and/or  expansion or compression of the trap ($\dot q_0$ term).
Geometrically the center of the ellipses 
in Fig. \ref{ellipse_fig} can move in the $\{q_1,q_2\}$ plane, and the size may change as well, 
but the orientation remains constant in time.
%
%
%
%
%
\subsection{Separation of two trapped interacting ions}
\begin{figure}
\includegraphics[width=7cm]{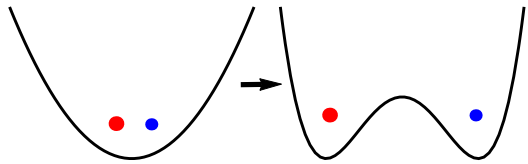}
\caption{(Color online) Scheme of the separation of two ions, from an external harmonic potential to a double well.}
\label{splitting_fig}
\end{figure}

%
Consider now the problem of separating (or recombining)  two interacting trapped ions as  in \cite{PMPRM2015}, see Fig. \ref{splitting_fig}.
The Hamiltonian of a system of two ions of masses $m_1$ and $m_2$ and charge $e$
located at $q_1>q_2$ in the laboratory frame is 
\begin{eqnarray}
 H&=&\frac{p_1^2}{2m_1}+\frac{p_2^2}{2m_2}+U,
 \nonumber\\
  U&=&\alpha(t)(q_1^2+q_2^2)+\beta(t)(q_1^4+q_2^4)+ \frac{C_c}{q_1-q_2},\nonumber
 \label{H_splitting_lab}
 \end{eqnarray}
 where $\alpha(t)$ and $\beta(t)$ are time dependent functions  \cite{PMPRM2015}. Typically $\alpha(0)>0$, $\beta(0)=0$, whereas at final time 
$\beta(t_f)>0$, $\alpha(t_f)<0$  to implement an evolution from a harmonic trap to a double well,  
see Fig. \ref{splitting_fig}.

To set a quadratic (approximate) Hamiltonian we proceed as in the previous sub-section:
first, we obtain the equilibrium positions $q_i^{(0)}$ of each ion by minimizing the potential $U$ 
and then expand the potential $U$ to second order in $q_i-q_i^{(0)}$.
If the equilibrium positions are denoted by $q_1^{(0)}=q_0/2$ and $q_2^{(0)}=-q_0/2$ (with $q_0$ being the equilibrium distance between ions)
this procedure gives the  quadratic Hamiltonian
\begin{eqnarray}
 H&=&\frac{p_1^2}{2m_1}+\frac{p_2^2}{2m_2} \nonumber\\
 &+&\frac{1}{2}\left(q_1-q_1^{(0)},q_2-q_2^{(0)}\right)
 K \left(  \begin{matrix}  q_1-q_1^{(0)}\\q_2-q_2^{(0)}  \end{matrix}\right),\nonumber
 \end{eqnarray}
with a $K$ matrix given by 
\begin{eqnarray}
k_1&=&k_2=2\alpha +3\beta q_0^2=k_0,\nonumber\\
k&=&\frac{2C_c}{q_0^3},\nonumber
\end{eqnarray}
%
%
and where the  equilibrium distance between ions $q_0=q_1^{(0)}-q_2^{(0)}$ is the solution of the quintic equation \cite{PMPRM2015,HS2006}
 \begin{equation}
 \label{quintic}
  \beta q_0^5+2\alpha q_0^3-2C_c=0.
\end{equation}
In this case 
\begin{eqnarray}
\label{tan_2theta_splitting}
\tan 2\theta
&=&\frac{\sqrt{m_1m_2}}{m_1-m_2}\left(\frac{4C_c}{2\alpha q_0^3+3\beta q_0^5+2C_c}\right),
\end{eqnarray}
which, in general, depends on time.
In principle, there are two ways to end up with a time-independent $\theta$ 
for time dependent   $\alpha$ and $\beta$:
\begin{itemize}
\item Equal masses, $m_1=m_2$, for which $\theta=\pm\pi/4$  regardless
of the time dependence of $\alpha$ and $\beta$ \cite{PMPRM2015}.
 
\item $\alpha$ and $\beta$ linked by   
%
%
\begin{equation}
\label{const}
\frac{\beta^3}{\alpha^5}=\textrm{constant},
\end{equation}
%
%
regardless of the masses. 
This implies that the products $\beta q_0^5$ and $\alpha q_0^3$ are constants.  
The particular case  $\beta=0$ corresponds to the one 
considered in the previous subsection (transport and expansions). The case $\beta\ne 0$ is interesting 
as it allows to separate or approach the two ions by decreasing or increasing $\alpha<0$
and $\beta>0$ according to  
Eq. (\ref{const}), i.e.,  in a double well confining  potential throughout the process.   
\end{itemize}
%
%
%
%
%
%
\subsection{Phase gates}
A phase gate can be implemented by applying  well designed time-dependent forces that depend on the internal states of the two ions in a linear trap \cite{phaseg}.  
The external harmonic trap for each ion has a fixed spring constant $k_0$.  
For a particular spin configuration the Hamiltonian becomes
\begin{eqnarray}
\label{Hphase}
H&=&\frac{p_1^2}{2m_1}+\frac{p_2^2}{2m_2}+\frac{1}{2}k_0(q_1^2+q_2^2)+\frac{C_c}{q_1-q_2}
\nonumber\\
&+&F_1(t) q_1+F_2(t) q_2.
\end{eqnarray}
Equilibrium positions and the equilibrium distance between the ions are
\begin{eqnarray}
q_1^{(0)}&=&\frac{B-2k_0^2\Delta(F_2+2F_1)}{6k_0^3\Delta}, \nonumber\\
q_2^{(0)}&=&\frac{-B-2k_0^2\Delta(F_1+2F_2)}{6k_0^3\Delta}, \nonumber\\
q_0&=&q_1^{(0)}-q_2^{(0)}=\frac{2B-2k_0^2\Delta(F_1+F_2)}{6k_0^3\Delta}, \nonumber
\label{q0}
\end{eqnarray}
where
\begin{eqnarray}
B&=&(F_1-F_2)^2k_0^4+\Delta^2, \nonumber\\
\Delta&=&\bigg\{-(F_1-F_2)^3k_0^6+27C_c k_0^8 \nonumber\\
&+&3\sqrt{3C_ck_0^{14}[-2(F_1-F_2)^3+27C_ck_0^2]}\bigg\}^{1/3}.\nonumber
\label{BD}
\end{eqnarray}
This leads to a $K$ matrix with
\begin{equation}
k=\frac{2C_c}{q_0^3},\;\;\;\;\;k_1=k_2=k_0, \nonumber
\end{equation}
so that
\begin{equation}
\tan 2\theta=\frac{\sqrt{m_1m_2}}{m_1-m_2}\, \frac{4C_c}{k_0q_0^3+2C_c}.  \nonumber
\end{equation}
The  angle $\theta$ is in general time-dependent, but it becomes constant in some cases, specifically for $m_1=m_2$,  and also for $F_1=F_2$. This latter
case in fact reduces to the transport of two ions considered before.
For different masses and forces  the ellipsoid rotates so that the modes $Q_1,\,Q_2$ are coupled.
A way out, if the forces are small so that the linear term in Eq. (\ref{Hphase}) may be considered a perturbation, is to define the modes for 
\cite{Juanjo}
\begin{equation}
H_0=\frac{p_1^2}{2m_1}+\frac{p_2^2}{2m_2}+\frac{1}{2}k_0(q_1^2+q_2^2)+\frac{C_c}{q_1-q_2}. \nonumber
\end{equation}  
These zeroth order modes, $Q_1(F_j=0),\,Q_2(F_j=0)$, $j=1,2$, are of course uncoupled since all coefficients in $H_0$ are time independent.
$H_0$ may thus be easily diagonalized. 
Specifically, all the $K$ matrix coefficients become equal, $\tan 2\theta=\frac{\sqrt{m_1m_2}}{m_1-m_2}$, 
and the equilibrium positions simplify to the constant values 
$\pm[\frac{C_c}{4k_0}]^{1/3}$.  
The inverse transformation  of Eq. (\ref{full_transf}) for these zeroth order modes,
 \begin{eqnarray}
  \label{full_transf_inv}
 \left(  \begin{matrix}  q_1\\q_2   \end{matrix}\right)=
 \left(  \begin{matrix}  q_1^{(0)}\\q_2^{(0)}\end{matrix}\right)_{\!\!F_j=0}+A_{F_j=0}^{\!-1}\!\left(  \begin{matrix}  Q_1\\Q_2\end{matrix}\right)_{\!\!F_j=0}, \nonumber
 \end{eqnarray}
enables us to write the perturbative linear terms in $H$ in terms of the uncoupled modes, so that $H$ is approximated as a sum of two uncoupled 
Hamiltonians \cite{Juanjo,phaseg}.  
%
%
%
%
%
\subsection{Anisotropic harmonic oscillator in 2D with rotation}
\label{anisotropic_osc_sec}
Consider now a single particle of mass $m$ trapped in a 2D anisotropic harmonic oscillator which is rotating around the $z$ axis with angular velocity $\dot \varphi$ \cite{Masuda}. 
Let us denote by $q_1=x$ and $q_2=y$ the lab frame coordinates of the ion, and by 
\begin{eqnarray}
 \tilde q_1(t)&=&q_1\cos\varphi(t)+q_2\sin\varphi(t),
\nonumber \\
 \tilde q_2(t)&=&-q_1\sin\varphi(t)+q_2\cos\varphi(t).
\label{rotated_q}
\end{eqnarray}
the coordinates in the rotating frame. The Hamiltonian in the lab frame is given by
\begin{eqnarray}
 H&=&\frac{p_1^2}{2m}+\frac{p_2^2}{2m}+\frac{1}{2}m\sum_{i=1}^2\omega_i^2 \tilde q_i(q_1,q_2;t)^2,
 \nonumber
 \end{eqnarray}
where $\omega_1\ne\omega_2$ are the angular frequencies along the rotating principal axes.  
(Note that for the trivial isotropic case $\omega_1=\omega_2$ the Hamiltonian is already uncoupled in the laboratory frame.)
$H$ can be written as 
\begin{equation}
 H=\frac{p_1^2}{2m}+\frac{p_2^2}{2m}+\frac{1}{2}(q_1,q_2)K\left(  \begin{matrix}  q_1\\q_2\end{matrix}\right),
\end{equation}
with the $K$ matrix given by
\begin{eqnarray}
 k_1&=&m\left(\omega_1^2\cos^2\varphi+\omega_2^2\sin^2\varphi\right)-k, \nonumber \\
 k_2&=&m\left(\omega_1^2\sin^2\varphi+\omega_2^2\cos^2\varphi\right)-k, \nonumber \\
 k&=&-\frac{m}{2}(\omega_1^2-\omega_2^2)\sin 2\varphi. \nonumber
\end{eqnarray}
%
%
 %
 Unlike the previous operations, there is no need to make the harmonic approximation around the equilibrium since $H$ is already quadratic.
 Moreover $q_1^{(0)}=q_2^{(0)}=0$ and, quite simply,  $\theta=\varphi$. The Hamiltonian $\tilde{H}$ takes the form in 
 Eq. (\ref{H_tilde_general}) without the linear term and with  $\Omega_i=\omega_i$, and $Q_i=\sqrt{m} \widetilde{q}_i$, i.e.,   
 the normal mode coordinates are the (mass-weighted) rotating coordinates  in Eq. (\ref{rotated_q}),
coupled  by the angular momentum term $-\dot{\theta}L_z$. 
We conclude that the 2D anisotropic problem is not separable by means of a linear point transformation of coordinates.

Making use of an additional physical interaction, it is possible to cancel the coupling term so that the resulting Hamiltonian is diagonal.
Specifically, if the particle is an ion of (positive) charge $e$, 
a homogeneous magnetic field $-B\hat{z}$ introduces, in the rotating frame, 
the diamagnetic and paramagnetic terms, see e.g. \cite{hami}, 
\begin{equation}
\frac{1}{2} \omega_L^2 (Q_1^2+Q_2^2)+\omega_L L_z, \nonumber
\end{equation}
where the Larmor frequency is $\omega_L=eB/(2m)$. 
(This term is invariant 
with respect to rotations around $\hat{z}$, so that it has the same form in laboratory coordinates and momenta.) 
Adjusting the magnetic field to exactly cancel $-\dot{\theta}L_z$ with $\omega_L=\dot{\theta}$, provides an uncoupled normal-mode Hamiltonian
with time dependent frequencies $[\omega_i^2+\omega_L^2]^{1/2}$.  
Note that the harmonic 2D potential complemented by a confining term in $z$-direction cannot be purely electrostatic as it would not obey Laplace's equation \cite{Masuda}. It could however be created by other means, 
for example as an effective pondermotive potential.

%
%
%
\subsubsection{Compressions and expansions}

\begin{figure}
\includegraphics[width=7cm]{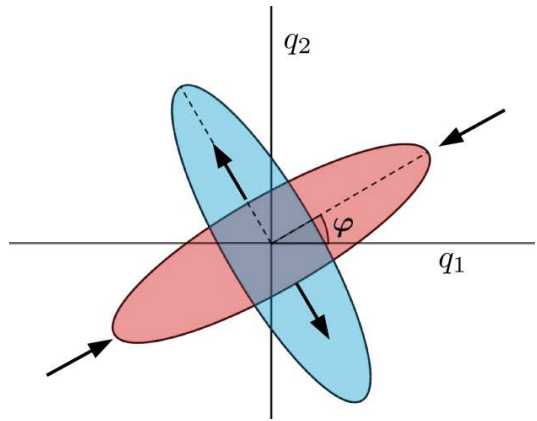}
\caption{(Color online) Compression and expansion along the non rotating principal axes of a 2D anisotropic oscillator amounts to a $\pi/2$ rotation of the potential}
\label{rotation_fig}
\end{figure}
We have pointed out before that $k=0$,
with $k_1$ and $k_2$ time-dependent leads to a constant $\theta$ and independent modes. For a single particle in a harmonic potential, 
this corresponds to time-dependent frequencies (expansions and compressions) along the nonrotating principal axes of the potential.    
Specific orthogonal compressions and expansions where the two normal mode frequencies interchange, amount at final time to a  $\frac{\pi}{2}$-rotation
of the potential,  see Fig. \ref{rotation_fig}, although the process itself is different.   
Slow adiabatic expansion and compression processes would connect initial and final excited energy levels differently from a true rotation, 
which may be important for inverse engineering operations.  
In the slow frequency manipulation the levels cross,  
so that their energy ordering changes. Take for example the initial states with vibrational quantum numbers 01 and 10 for the principal directions 1 and 2 and such that $\omega_1<\omega_2$.  Then the energies satisfy initially $E_{01}>E_{10}$.
If the values of the frequencies are interchanged along the process the energies also switch, $E_{10}>E_{10}$. 
On the contrary, a slow true rotation does not produce crossings 
and energy reordering.     

\section{Discussion}
Motivated by the need to inverse engineer the dynamics of trapped ions and other systems in the small oscillations regime, 
we have studied the possibility to define, via linear point transformations,  independent dynamical normal modes for
two dimensional systems under
time-dependent external control. Whereas the analysis of further dimensions is certainly worthwhile, 
note the physical relevance of two dimensions,  as they suffice to describe pairs of ions in linear traps, and  
universal quantum computing may be achieved by combining operations on one and two qubits.  We also expect that the results found here may set a useful guide for further dimensions.  
The condition that determines the coupling of the
modes turns out to be the rotation of the harmonic potential in the (laboratory) 2D coordinate space. Nonrotating 
potentials lead to uncoupled dynamical modes. Different examples have been analyzed and in some of them ways to avoid the coupling have been pointed out:  by a specific design of the time dependence of the control parameters in separation operations,  by
adding  compensating terms in the Hamiltonian in rotations, or perturbatively in phase gates. 
Point transformations are the ones used for time-independent normal-mode analysis, so they are a natural choice. Moreover  they are easy 
to understand, visualize, and implement.   
More general (mixed) canonical transformations have not been considered in this paper, but they are in principle possible  \cite{Inva,Manko,Ji,Xu,AbdallaLeach,Menouar}
and will be discussed elsewhere.    
Generically their physical meaning, definition, and practical 
use become  more involved, so only simplified potential configurations and dynamics are typically 
worked out explicitly \cite{Inva,Manko,Ji,Xu,AbdallaLeach,Menouar}.

\begin{acknowledgments}
We acknowledge support by MINECO (Grant FIS2015-67161-P) and the program UFI 11/55 of UPV/EHU.
M.P. acknowledges a fellowship by UPV/EHU.
\end{acknowledgments}
\appendix

\section{Derivation of the Hamiltonian of the spring system}
\label{spring_system_appendix}
\begin{figure}
\includegraphics[width=7cm]{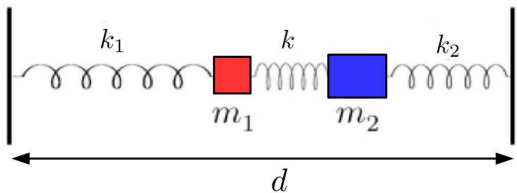}
\caption{(Color online) Mechanical system of two masses connected to each other and to the walls 
by springs with time-dependent spring ``constants''. This system is found to be mathematically equivalent to many 
of the trapped ions systems considered throughout the text.}
\label{springs_fig}
\end{figure}
In this Appendix we find  the Hamiltonian describing the dynamics of two masses connected by springs as illustrated in   
Fig. \ref{springs_fig}. The three springs are assumed to have zero natural length but time-dependent spring ``constants''.
If $q_i$ is the lab frame coordinate of $m_i$ measured from the fixed left wall,
the Hamiltonian is given by
\begin{eqnarray}
 H&=&\frac{p_1^2}{2m_1}+\frac{p_2^2}{2m_2}+U \nonumber\\
 U&=&\frac{1}{2}k_1q_1^2+\frac{1}{2}k_2\left(d-q_2\right)^2+\frac{1}{2}k (q_2-q_1)^2, \nonumber
\end{eqnarray}
where $p_i$ is the conjugate momentum of the coordinate $q_i$.
By solving the set of equations $\partial_{q_i} U=0$ for $i=1,2$ we find the equilibrium positions of the two connected masses at
%
%
\begin{eqnarray}
 q_1^{(0)}&=&q_0\left(\frac{k}{k_1}\right), \nonumber\\
 q_2^{(0)}&=&q_1^{(0)}+q_0, \nonumber
\end{eqnarray}
where $q_0=q_2^{(0)}-q_1^{(0)}$, the equilibrium distance between masses, is given by 
\begin{eqnarray}
 q_0&=&d\left[\frac{k_1k_2}{k_1k_2+k(k_1+k_2)}\right]. \nonumber
\end{eqnarray}
The equilibrium positions are generally moving, 
they depend on time because of the time dependence of $k,\,k_1$, and $k_2$. 
We can now expand the coupling potential $U$ around its equilibrium position (small oscillations), and up to a purely time dependent function, 
we have
\begin{eqnarray}
 U&=&\frac{1}{2}\!\left(q_1\!-\!q_1^{(0)},q_2\!-\!q_2^{(0)}\right)\left(  \begin{matrix}  k_1+k&-k\\-k&k_2+k  \end{matrix}\right)
  \left(  \begin{matrix}  q_1\!-\!q_1^{(0)}\\q_2\!-\!q_2^{(0)}  \end{matrix}\right),
 \nonumber
\end{eqnarray}
where  the $q_i-q_i^{(0)}$  measure the displacement of mass $m_i$ from its (moving) equilibrium position. 
The full Hamiltonian then may be written exactly as in Eq. (\ref{starting_H}).  
%
%

\section{Quantum treatment}
\label{quantum_appendix}
The results in the main text regarding the form of the Hamiltonians and transformations can be used directly 
in quantum mechanical systems.
The starting point is  the 2D time dependent Schr\"odinger equation
\begin{eqnarray}
 i\hbar \partial_t  \psi(q_1,q_2;t)=H(q_1,q_2;t) \psi(q_1,q_2;t),
 \label{seq}
\end{eqnarray}
with $H(q_1,q_2;t)$ given in Eq. (\ref{starting_Hqp}).
Let us now consider the use of the new coordinates (\ref{A_change_coordinates}) and define
\begin{equation}
\Psi=\Psi(Q_1,Q_2;t)\equiv\psi[q_1(Q_1,Q_2;t),q_2(Q_1,Q_2;t);t].  
\end{equation}
We now calculate the time derivative of the transformed wavefunction taking into account the time dependences separately and applying the chain rule,  
\begin{eqnarray}
i\hbar\partial_t\Psi&=&i\hbar\partial_t\psi[q_1(Q_1,Q_2;t),q_2(Q_1,Q_2;t);t]\nonumber\\
&=&i\hbar\left[
\partial_t
+\partial_t q_1\partial_{q_1}
+\partial_t q_2\partial_{q_2}
\right]\Psi\nonumber\\
&=&\left[H(Q_1,Q_2)-(p_1,p_2)\left(\begin{matrix} \partial_t q_1\\\partial_t q_2\end{matrix}\right)\right]\Psi,
\label{seq2}
\end{eqnarray}
with $p_j=-i\hbar\partial_{q_j}$.
The first term is just the transformed Hamiltonian, i. e., the original Hamiltonian written in the new coordinates with the usual definition
$P_j=-i\hbar\partial_{Q_j}$, 
and the second term is an inertial contribution due to the time dependence of the transformation.
It is now clear that the effective Hamiltonian is
\begin{eqnarray}
 \widetilde H(Q_1,Q_2)
 &=&H(Q_1,Q_2)-(P_1,P_2)A\left(\begin{matrix} \partial_t q_1\\\partial_t q_2\end{matrix}\right),
  \label{H_tilde_quantum}
\end{eqnarray}
where relation (\ref{pP}) has been used to write the old momenta in terms of the new ones.
The explicit time derivative of the $q_i$ coordinates in (\ref{H_tilde_quantum})  can be calculated directly by inverting transformation (\ref{A_change_coordinates}), 
\begin{eqnarray}
 \left(\begin{matrix} \partial_t q_1\\  \partial_t q_2\end{matrix}\right)=\left(\begin{matrix} \dot q_1^{(0)}\\  \dot q_2^{(0)}\end{matrix}\right)
 +\partial_t (A^{-1})\left(\begin{matrix} Q_1\\  Q_2\end{matrix}\right),
\end{eqnarray}
which leads finally to an effective Hamiltonian
\begin{eqnarray}
 \widetilde H&=&H
 -(P_1,P_2) A \left(\begin{matrix} \dot q_1^{(0)}\\ \dot q_2^{(0)}\end{matrix}\right)
 -(P_1,P_2) A \partial_t (A^{-1})\left(\begin{matrix} Q_1\\  Q_2\end{matrix}\right)\nonumber\\
 &=&H-(P_1,P_2) A \left(\begin{matrix} \dot q_1^{(0)}\\ \dot q_2^{(0)}\end{matrix}\right)
 -\dot\theta \left(Q_1P_2-Q_2P_1\right),  
\end{eqnarray}
where, for writing the last term, we have used the relation
\begin{equation}
  A \partial_t (A^{-1})=\dot\theta \left(\begin{matrix} 0 &-1\\1&0\end{matrix}\right).
\end{equation}
The effective Hamiltonian when using transformed coordinates thus
takes exactly the same form as in classical mechanics, Eq. (\ref{H_tilde_general}).


\end{document}